\documentclass[aps,onecolumn,showpacs,preprintnumbers,amsmath,amssymb]{revtex4-1}
\usepackage{amsmath,amssymb,graphics}
\usepackage{graphicx}
\usepackage{subfigure}
\usepackage{colortbl}
\usepackage{color}

\begin{document}

\title{Localization of Gravitino Field on Branes}

\author{Yun-Zhi Du$^{1,2}$\footnote{duyzh13@lzu.edu.cn},
        Li Zhao$^2$\footnote{{lizhao@lzu.edu.cn}},
        Xiang-Nan Zhou$^3$\footnote{zhouxn10@lzu.edu.cn},
        Yi Zhong$^2$\footnote{zhongyi11@lzu.edu.cn},
        and Yu-Xiao Liu$^2$\footnote{liuyx@lzu.edu.cn, corresponding author}}

\affiliation{
{\small$^1$Institute of Theoretical Physics,
  Datong University, Datong 037009, P.R.China},\\
{\small $^2$Institute of Theoretical Physics,
  Lanzhou University, Lanzhou 730000, P.R.China},\\
{\small$^3$College of Physics and Information Engineering,
  Shanxi Normal University},\\
{\small Linfen 041004, P.R.China}
  }

\begin{abstract}
In this paper, we investigate the localization of a bulk gravitino field on the scalar-tensor branes and compare the result with that in the Randall-Sundrum-1 (RS1) model. The coupled chiral equations for the Kaluza-Klein (KK) modes of the gravitino field are obtained by fixing the gauge $\Psi_5=0$ and using the chiral KK decompositions. It is shown that, in the RS1 model for the left- and right-handed zero modes of the gravitino field, only one of them can be localized near one brane. For the massive modes, both chiral modes survive and the lower KK modes are localized near the IR brane from the four-dimensional physical coordinate point of view. However, for the scalar-tensor brane model, the localization of the gravitino chiral zero modes depends on the coupling parameter $\lambda$, and they will be not localized around anyone brane within a certain range of the parameter $\lambda$, which is quite different from the RS1 model. Furthermore, we also give the corresponding mass spectra of the massive KK gravitinos in the scalar-tensor model.
\end{abstract} 

\maketitle

\section{Introduction}\label{sec1}

It is well-known that our visible world consists of one-dimensional time and three-dimensional space. However, the Kaluza-Klein (KK) theory \cite{Kaluza1921,Klein1926} and string theory indicate that there are extra spacial dimensions beyond our familiar ones. Inspired by string theory, Antoniadis, Arkani-Hamed, Dimopoulos, and Dvali suggested that our four-dimensional universe is a membrane embedded in a higher-dimensional spacetime with flat large extra dimensions \cite{Arkani-Hamed1998,Antoniadis1998}. Subsequently, Randall and Sundrum (RS) proposed a warped braneworld scenario in a five-dimensional Anti-de Sitter (AdS) spacetime, where the extra dimension is warped \cite{Randall1999,Randall1999a}. The braneworld theory has attracted much attention because it can provide an alternative mechanism to solve the hierarchy problem (the 16 order difference between the weak scale and the Planck scale).
Furthermore, extra dimensions in such theory may be probed by current experiments \cite{Antoniadis1999,Abbott2001,Nussinov2002,Pankov2005,Abulencia2006,Abdallah2009,Franceschini2011,Basto-Gonzalez2013,Sun2014}.
This is different from the original KK theory and string theory, where extra dimensions are very hard to be detected by current or near future experiments. Some of the possible experimental probes of extra spatial dimensions include the possible creation of mini-black holes \cite{Cheung2002,Casanova2006,Rizzo2007,Casadio2012} and the collider cross~ sections for interactions between the Standard Model particles and KK particles~\cite{Park2002,Li2013,Sahin2015} at the Large Hadron Collider as well as deviations of Newton's law of gravity at the micro-meter or smaller distance in current table top experimental searches~\cite{Hoyle2001,Hoyle2004,Adelberger2007,Kapner2007,Shao2015}.

In the braneworld scenario, all particles/fields in the Standard Model should be localized on a $(3+1)-$dimensional brane to recover the four-dimensional lower energy field theory. In Refs.~\cite{Gherghetta2011,Grossman2000,Gherghetta2000,Huber2001,Huber2003}, the authors used the freedom of the mass terms of bulk scalar and fermion fields to construct a bulk Standard Model. To explain the fermion mass (flavor) hierarchy in the four-dimensional effective theory, all fields or matters should be in the bulk, while the Higgs field is confined near the IR brane in the Randall-Sundrum-1 (RS1) model. In the recent work \cite{Agashe2015}, it was shown that the Higgs field can also be in the bulk. Therefore, we need localization mechanisms for different spin fields in the bulk to make our world appears effectively $(3+1)-$dimensional up to some energy scale. Localization of various matters in some kinds of braneworld models has been extensively investigated for examples in Refs. \cite{Gherghetta2011,Csaki2000,Gremm2000,Bajc2000,Shiromizu2000,Oda2001,Mukhopadhyaya2004,Chatterjee2005,Melfo2006,Liu2007a,SouzaDutra2008,Tahim2009,Almeida2009,Bazeia2009,Guerrero2010,Landim2011,Yang2012,Fu2012,Cruz2013,Xie2015,Zhao2015}. In this paper, we are interested in the localization of a five-dimensional massive gravitino field on thin branes.

Gravitino is the particle with spin $3/2$. It was proposed as the supersymmetric partner of graviton in the supergravity theory. In particular, this particle is regarded as a candidate of dark matter in cosmology \cite{Chun1994,Viel2005,Steffen2006,Panotopoulos2007,Ding2012,Savvidy2013}. It was shown that if the axino mass is around $1$ MeV and gravitino mass around $1$ eV, then hot axions and gravitinos could act like hot dark matters~\cite{Chun1994}. In Ref.~\cite{Steffen2006}, gravitino is a promising candidate for the cold dark matter with new bounds on its mass. However, the light gravitinos with mass in 1 eV to 1 MeV range as the dark matter candidate are muddled by early-universe uncertainties~\cite{Feng2010}. The authors in Ref.~\cite{Feng2010} had explained how upcoming data from colliders may clarify this uncertainty picture. Furthermore, in the supersymmetric Standard Models the lightest supersymmetric particle is an ultralight gravitino ($m_{3/2} \sim $ 1 eV). A simple test for the lightness of gravitino at the Large Hadron Collider was proposed in Ref.~\cite{Shirai2009}. Recently, some attention was paid to gravitino fields in black hole backgrounds~\cite{Arnold2014,Piedra2011,Yale2009,Khlopov2006}. Therefore, the localization and mass spectrum of a gravitino field on a $(3+1)-$dimensional brane are of great significance in the brane scenario.

The localization of a gravitino field ($\Psi_M$) was investigated in Refs.~\cite{Bajc2000,Oda2001,Liu2007a,Gherghetta2001,Oda2001a,Lee2008,Zhou2017}. Just like the case of a free Dirac fermion, a free gravitino cannot be confined to  a RS-like brane in a five-dimensional spacetime. Thus, the authors of Refs.~\cite{Oda2001,Gherghetta2001} studied the localization of zero mode of a five-dimensional bulk gravitino field by introducing a bulk mass term. And the interaction between the gravitino and the function $f(\phi)$ of the background scalar field $\phi$ has been considered to investigate the KK modes of a five-dimensional gravitino field in the $f(R)$ thick brane models \cite{Zhou2017}. It was shown in Refs. \cite{Liu2007a,Oda2001a} that, by taking some gauges such as $\Psi_5=0$, the zero mode of an interacting bulk gravitino in a $D-$dimensional spacetime ($D\geq6$) has similar localization property as that of an interacting bulk Dirac fermion. In a six-dimensional gauged supergravity model, the brane-induced gravitino mass term would lead to a power-like suppressed on the effective four-dimensional gravitino mass \cite{Lee2008}. Furthermore, the coupling of the gravitino KK modes with other fermions and with their scalar supersymmetrical partners on the brane was investigated in Ref.~\cite{Hewett2004}. In this manuscript, we will study the KK modes of the five-dimensional gravitino field on the thin branes.

The most typical thin brane model in the braneworld senario is the RS1, in which there are two thin branes (positive tension brane (UV) and negative one (IR)). To solve the gauge hierarchy problem, the extra dimension is warped, and our four-dimensional world should be resided on the negative tension brane in this model. However our four-dimensional expanding universe on the negative tension brane can not be recovered \cite{Shiromizu2000,Csaki1999,Cline1999}, which is a severe cosmological problem in the RS1 model. It can be addressed by confining our world on the positive tension brane in the scalar-tensor brane model \cite{Yang2012}. Furthermore, the localization of various five-dimensional bulk fields on the scalar-tensor branes has been investigated in Refs. \cite{Yang2012,Xie2015}. The infinite massive KK modes of gravity were obtained, and the mass gap was tiny, i.e., $10^{-4}$eV. However, there is no violation of the gravity effect on LHC, because the coupling to matter fields on the positive tension brane is very weak. In this paper, we would like to consider a five-dimensional massive gravitino field in the scalar-tensor brane model \cite{Yang2012} by taking the gauge $\Psi_5=0$. Our aim is to obtain the localization of the chiral zero modes and the mass spectrum of the KK gravitinos, and compare the localization property of this field with that in the RS1 model.

This work is organized as follows. In Sec. \ref{sec2}, we consider a massive gravitino field in a five-dimensional spacetime with the extra dimension compactified on $S^1/Z_2$. In particular, the coupled chiral equations and the orthonormality conditions of the left- and right-handed KK gravitinos are derived. Then we will review the localization of the gravitino field on the RS1 brane, and give the mass spectrum of the KK modes. Furthermore, several lower massive KK modes of the gravitino will be showed in Sec. \ref{sec3}.  In Sec. \ref{sec4}, the localization and mass spectrum of the gravitino field on the scalar-tensor branes are investigated, and the results are compared with that in the RS1 model. Finally, the discussion and conclusion are presented in Sec. \ref{secConclusion}.

\section{Localization of a gravitino field on thin branes}
\label{sec2}

In this section, we consider the localization of a gravitino field on a thin brane embedded in a five-dimensional spacetime with coordinates $x^M=(x^\mu,y)$. Here, the five-dimensional spacetime indices are denoted by $M,N,\cdots=0,1,2,3,5$ and the four-dimensional Minkowskian spacetime indices $\mu,\nu,\cdots=0,1,2,3$. To avoid deviation from the detection by current experiments, one can think that the extra dimension $y$ is finite and the simplest way is to assume a periodic geometry of the extra dimension such as a circle $S^1$ with radius $R$. However, the chiral fermions in the Standard Model cannot be described by compactifying bulk fermions on a circle. Instead the fifth dimension can be compactified on a line segment $S^1/Z_2$ with $Z_2$ representing the identification $y\leftrightarrow-y$ (so we have $0\leq y\leq\bar y=\pi R$). This is the well-known $Z_2$ orbifold symmetry \cite{Randall1999a,Yang2012}. There are two thin 3-branes located at two orbifold fixed points $y=0$ and $y=\bar y$.

The line element of the five-dimensional spacetime is given by
\begin{eqnarray}
ds^2=e^{-2\sigma}\eta_{\mu\nu}dx^\mu dx^\nu+dy^2,\label{LE}
\end{eqnarray}
where the warped factor $e^{-2\sigma}$ is only the function of $|y|$. Based on the metric (\ref{LE}), the warped five-dimensional vielbeins $e_M^{~~\bar M}$ and the flat four-dimensional ones $\hat e_\mu^{~~\bar\mu}$ read
\begin{eqnarray}
e_M^{~~\bar M}= \left[
\begin{array}{cc}
    e^{-\sigma}~\hat e_\mu^{~~\bar\mu}&0\\0&1
\end{array}\right],~~
e^M_{~~\bar M}=\left[
\begin{array}{cc}
    e^{\sigma}~\hat e^\mu_{~~\bar\mu}&0\\0&1
\end{array}\right], ~~
\hat e_\mu^{~~\bar\mu}=\left[
                         \begin{array}{cc}
                           ~  I &  ~ \\
                           ~    &  I  \\
                         \end{array}
                       \right]
.\nonumber
\label{VE1}
\end{eqnarray}
Here, $\bar{M}$ and $\bar{\mu}$ denote the (4+1)-dimensional and (3+1)-dimensional local Lorentz indices, respectively.
With the relations $e_{M\bar M}=g_{MN}e^{N}_{~~\bar M}$ and $e^{M\bar M}=g^{MN}e_{N}^{~~\bar M}$, we have
\begin{eqnarray}
e_{M\bar M}\!=\!\!\left[
\begin{array}{cc}
\!\!e^{-\sigma}\hat e_{\mu\bar\mu}& \!\!0\\
0&1
\end{array}\!\!\right],~~
e^{M\bar M}\!=\!\!\left[
\begin{array}{cc}
\!\!e^{\sigma}\hat e^{\mu\bar\mu}&\!\!0\\
0&1
\end{array}\!\!\right],~~
\hat e_{\mu\bar\mu}
    \!=\!\eta_{\bar\mu\bar\nu}\hat e_\mu^{~\bar\nu}.
    \label{VE2}
\end{eqnarray}

In Refs.~\cite{Gherghetta2000,Bergshoeff2000,Altendorfer2001,Alonso2000,Falkowski2000}, the authors gave the complete supergravity action in the RS1 model, which includes the gravitino and graviphoton together with the graviton. In an $AdS_5$ spacetime, the graviphoton is set to zero \cite{Gherghetta2000}. In order to study the supersymmetry breaking, the additional gravitino kinetic and mass terms were considered in Ref. \cite{Gherghetta2001}.

It is known that the Dirac fermion mass parameter is odd under the $Z_2$ transformation in a five-dimensional spacetime with the extra spacial dimension compactified on $S^1/Z_1$, so is the gravitino mass parameter. Therefore, for the case of a gravitino field, the simplest additional mass term can be proportional to $\sigma'\equiv\partial_y \sigma$. Furthermore, the gravitino mass arises from an odd `kink' profile background scalar field with a vacuum expectation value and the vacuum configuration of the scalar field can be regarded as an infinitely thin domain wall. Hence, we adopt the action for a gravitino field with an additional mass term in the five-dimensional spacetime as~\cite{Gherghetta2001}
\begin{eqnarray}
S_{\frac{3}{2}}=\int d^5x\sqrt{-g}i\bar\Psi_M\left(\Gamma^{[MNL]}D_N
  -\frac{3\sigma'}{2}\Gamma^{[ML]}\right)\Psi_L,\label{S}
\end{eqnarray}
where $\Gamma^{[MNL]}=\Gamma^{[M}\Gamma^N\Gamma^{L]}$. The matrices $\Gamma^M$ provide a four-dimensional representation of the Dirac matrices ($\gamma^\mu$ and $\gamma^5$) with the chiral representation in a five-dimensional warped spacetime. The covariant derivation $D_N\Psi_L$ is defined as
\begin{eqnarray}
D_N\Psi_L \equiv \partial_N\Psi_L-\Gamma^M_{~~NL}\Psi_M+\omega_N\Psi_L
\end{eqnarray}
with the spin connection $\omega_N=\frac{1}{4}{\omega_{N}}^{\bar N\bar L}\Gamma_{\bar N}\Gamma_{\bar L}$. From the formulas (\ref{VE1}) and (\ref{VE2}), the relation between the warped gamma matrices $\Gamma^M$ and the Minkowskian ones $\Gamma^{\bar M}=(\Gamma^{\bar\mu},\Gamma^{\bar5})=(\gamma^{\bar\mu},\gamma^5)$ is given by $\Gamma^M=e^M_{\,\,\,\,\bar M}\Gamma^{\bar M}$, i.e.,
\begin{eqnarray}
\Gamma^\mu=e^\sigma\gamma^\mu,~~~~~\Gamma^5=\gamma^5.
\end{eqnarray}
The commutation relations of two gamma matrices are $[\Gamma^M,\Gamma^N]=2g^{MN}$ and $[\Gamma^{\bar{M}},\Gamma^{\bar{N}}]=2\eta^{{\bar{M}}{\bar{N}}}$, respectively. With the metric (\ref{LE}), we have $\omega_\mu=-\frac{1}{2}\sigma' e^{-\sigma}\gamma_\mu\gamma_5$ and $\omega_5=0$. Then the nonvanishing components of the covariant derivative $D_N\Psi_L$ can be calculated as follows:
\begin{eqnarray}
D_\mu\Psi_\nu
&=&\partial_\mu\Psi_\nu-\sigma'e^{-2\sigma}\eta_{\mu\nu}\Psi_5
-\frac{\sigma'}{2}e^{-\sigma}\gamma_\mu\gamma_5\Psi_\nu,\label{G1}\\
D_\mu\Psi_5
&=&\partial_\mu\Psi_5+\sigma'\Psi_\mu
   -\frac{\sigma'}{2}e^{-\sigma}\gamma_\mu\gamma_5\Psi_5,\label{G3}\\
D_5\Psi_\mu
&=&\partial_y\Psi_\mu+\sigma'\Psi_\mu,\label{G4}\\
D_5\Psi_5
&=&\partial_y\Psi_5+\sigma'e^{-2\sigma}\Psi_5.
\end{eqnarray}
The Rarita-Schwinger equation for the bulk gravitino field $\Psi_M$ from the action (\ref{S}) is
\begin{eqnarray}
\Gamma^{[M}\Gamma^N\Gamma^{L]}D_N\Psi_L=\frac{3\sigma'}{2}\Gamma^{[M}\Gamma^{L]}\Psi_L,\label{RS}
\end{eqnarray}
and the boundary terms will vanish provided that
\begin{eqnarray}
\left(\bar\Psi_{R\mu}\delta\Psi^\mu_{L}\right)\mid_{0,\bar y}
=\left(\bar\Psi_{L\mu}\delta\Psi_R^{\mu}\right)\mid_{0,\bar y}=0.\label{BCD}
\end{eqnarray}

Next, we try to give the coupling equations of the left- and right-handed KK modes for the gravitino field described in the action (\ref{S}) by directly taking the gauge $\Psi_5=0$ as did in Ref.~\cite{Oda2001}. Decomposing the five-dimensional fields $\Psi_\mu$ into four-dimensional Kaluza-Klein modes \cite{Gherghetta2001}
\begin{eqnarray}
\Psi_\mu(x,y)
&=&\sum_n\left[\psi^{(n)}_{L\mu}(x)\xi_{Ln}(y)
+\psi^{(n)}_{R\mu}(x)\xi_{Rn}(y)\right],\label{CD}
\end{eqnarray}
for $M=\lambda$, the equation (\ref{RS}) can be simplified to
\begin{equation}
\gamma^{[\lambda\mu\nu]}\partial_\mu\psi^{(n)}_{L\nu,R\nu}
=m_n\gamma^{[\lambda\nu]}\psi^{(n)}_{R\nu,L\nu},
\end{equation}
which is the four-dimensional massive Rarita-Schwinger equation for the spin $3/2$ field $\tilde\psi^{(n)}_{\nu}(x)$. And the KK modes $\xi_{Ln,Rn}(y)$ satisfy
\begin{eqnarray}
\Big[\pm\partial_y+(\frac{3}{2}\mp1)\sigma'\Big]\xi_{Ln,Rn}
=m_ne^{\sigma}\xi_{Rn,Ln},\label{CEOM}
\end{eqnarray}
which is consistent with the result in Ref. \cite{Gherghetta2001}. From above equations, we can obtain the solutions of the left- and right-handed zero modes:
\begin{eqnarray}
\xi_{L0}(y)\propto e^{-\frac{1}{2}\sigma},~~~~~~
\xi_{R0}(y)\propto e^{\frac{5}{2}\sigma}.\label{EOMzeros}
\end{eqnarray}

For the massive modes, we proceed with the coordinate and field transformations: $dy=e^{-\sigma}dz,~\xi_{Ln,Rn}(y)=e^\sigma\bar{\xi}_{Ln,Rn}(z)$. Then the chiral coupling equations (\ref{CEOM}) can be rewritten as
\begin{eqnarray}
\left(\pm\partial_z+\frac{3}{2}\partial_z\sigma\right)\bar\xi_{Ln,Rn}
=m_n\bar\xi_{Rn,Ln}.\label{CEOMz}
\end{eqnarray}
Thus, it is easy to see that KK modes $\bar{\xi}_{Ln,Rn}$ satisfy the following Schr\"{o}dinger-like equations:
\begin{eqnarray}
\left(-\partial_z^2+V_{L,R}(z)\right)\bar{\xi}_{Ln,Rn}=m_n^2\bar{\xi}_{Ln,Rn},\label{Seq}
\end{eqnarray}
where $m_n$ is the mass of the gravitino KK mode $\tilde{\psi}^{(n)}_{L\nu,R\nu}$ and the effective potentials read
\begin{eqnarray}
V_{L,R}(z)=\frac{9}{4}\left(\partial_z\sigma\right)^2\mp\frac{3}{2}\partial_z^2\sigma.\label{effectiveV}
\end{eqnarray}
The Schr\"{o}dinger-like equations (\ref{Seq}) can also be written as
\begin{eqnarray}
 Q^{\dagger} Q \bar{\xi}_{Ln}=m_n^2\bar{\xi}_{Ln}, ~~~~
 Q Q^{\dagger}  \bar{\xi}_{Rn}=m_n^2\bar{\xi}_{Rn}
\end{eqnarray}
with $Q=\partial_z+\frac{3}{2}\partial_z\sigma$, which indicates that there is no tachyon KK gravitino with negative mass square $m_n^2$. Finally, we need the orthonormality conditions

\begin{eqnarray}\label{orthogonality}
\int \!\!dy e^{-\sigma} \xi_{Ln}(y)\xi_{Lk}(y)&=&\int dz \bar\xi_{Ln}(z)\bar\xi_{Lk}(z)=\delta_{nk} ,   \\
\int\!\! dy e^{-\sigma} \xi_{Rn}(y)\xi_{Rk}(y)&=&\int dz \bar\xi_{Rn}(z)\bar\xi_{Rk}(z)=\delta_{nk} , ~~~~~~~   \\
\int\!\! dy e^{-\sigma} \xi_{Ln}(y)\xi_{Rk}(y)&=&\int dz \bar\xi_{Ln}(z)\bar\xi_{Rk}(z)=0
\end{eqnarray}
to obtain the effective four-dimensional action for the massless and massive gravitino fields:
\begin{eqnarray}
S_{\frac{3}{2}}
 &=&\sum_n\int d^4x\,i\Big(
     \bar{\psi}^{(n)}_{L\lambda}\gamma^{[\lambda\mu\nu]}
      \partial_\mu\psi_{L\nu}^{(n)}
   -m_n\bar{\psi}^{(n)}_{L\lambda}\gamma^{[\lambda\nu]}
        \psi_{R\nu}^{(n)} \nonumber \\
 && +\bar{\psi}^{(n)}_{R\lambda}\gamma^{[\lambda\mu\nu]}
    \partial_\mu\psi_{R\nu}^{(n)}
 -m_n\bar{\psi}^{(n)}_{R\lambda}\gamma^{[\lambda\nu]}
      \psi_{L\nu}^{(n)}
\Big)\nonumber\\
&=&\sum_n\int d^4x~ i\bar{\psi}^{(n)}_{\lambda}\Big(
    \gamma^{[\lambda\mu\nu]}
     \partial_\mu
  -m_n\gamma^{[\lambda\nu]}
   \Big)\psi_{\nu}^{(n)}.
\end{eqnarray}
Note that the orthonormality conditions (\ref{orthogonality}) are very essential to check whether the KK modes of the gravitino field can be localized on a brane.

Next we will firstly review the localization of the gravitino field in the RS1 model, and then investigate this field on the scalar-tensor brane and give its KK mass spectrum, which may be the signal of warped extra dimensions in future experiments.

\section{Review of the localization of a gravitino field on the RS1 brane}
\label{sec3}

In this part, we review the localization of the gravitino field on the RS1 branes in detail based on Ref. \cite{Gherghetta2001}. In the RS1 brane model, the five-dimensional fundamental scale is the Planck scale $M_{\textrm{pl}}\approx10^{16}\textrm{TeV}$ and the solution of the warped factor is
\begin{eqnarray}
\sigma(y)=k |y| \label{WarpedFactorRS2}
\end{eqnarray}
with the anti-de Sitter curvature $k \sim M_{\textrm{pl}}$. 

\begin{figure*}[htb]
\begin{center}
\includegraphics[width=6.5cm]{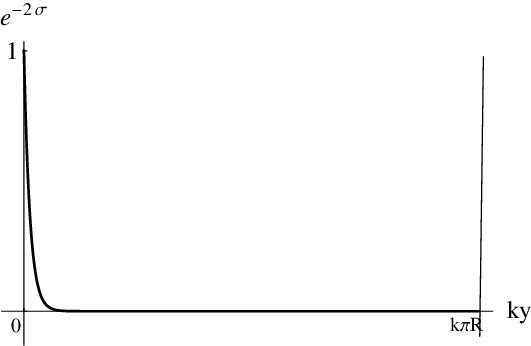}
\includegraphics[width=6.5cm]{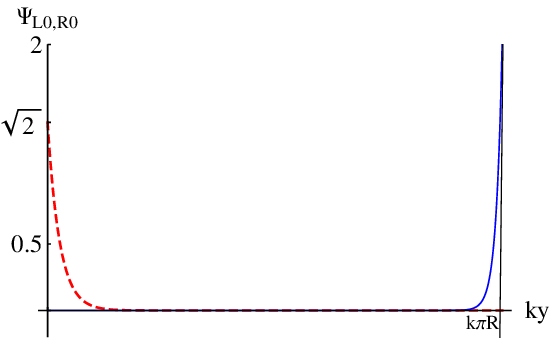}
\end{center}
\caption{\label{RS0} The shapes of the warped factor $e^{-2\sigma}$ and chiral zero modes $\Psi_{L0}\equiv e^{-\frac{\sigma}{2}}\frac{\xi_{L0}}{\sqrt{k}}$ (the red dashed line), $\Psi_{R0}\equiv e^{-\frac{\sigma}{2}}\frac{\xi_{R0}}{\sqrt{k}}$ (the blue thick line). }
\end{figure*}

With the solution (\ref{WarpedFactorRS2}), the left- and right-handed zero modes of the gravitino field (\ref{EOMzeros}) read
\begin{eqnarray}
\xi_{L0}=\frac{1}{\sqrt{N_{L0}}}e^{-\frac{1}{2}k\mid y\mid},~~~~~~
\xi_{R0}=\frac{1}{\sqrt{N_{R0}}}e^{\frac{5}{2}k\mid y\mid},
\end{eqnarray}
where $N_{L0}$ and $N_{R0}$ are the normalization constants. As the result of the orthogonality conditions (\ref{orthogonality}), the normalization constants $N_{L0}=\frac{1}{2k}(1-e^{-2k\pi R})\approx \frac{1}{2k}$, $N_{R0}=\frac{1}{4k}(e^{4k\pi R}-1) \approx\frac{1}{4k}e^{4k\pi R}$, where we have considered that $kR\approx12$ \cite{Randall1999}. From Fig. \ref{RS0}, it seems that the left-handed zero mode is localized on the brane at $y=0$ (positive tension (UV) brane), while the right-handed zero mode on the brane at $y=\pi R$ (negative tension (IR) brane). However, when considering the boundary conditions (\ref{BCD}), we need that either $\xi_{R0}$ is fixed on the boundaries with $\xi_{R0}|_{0,\pi R}=0$ or $\xi_{L0}$ is fixed. Thus we can obtain only
one of the left- and right-handed zero modes on a brane. In fact this is how the four-dimensional chirality is recovered from five-dimensional bluk fermion fields, and it is the result of compactifying fermions on the $S^1/Z_2$ extra dimension. This property will be important to describe the Standard Model fermions since the left- and right-handed fermions transform differently under the electroweak gauge group \cite{Gherghetta2011}.

For the massive KK modes, the effective potentials $V_{L,R}$ contain the second-order derivative of $\sigma$, which leads to the appearance of the delta functions. Thus, firstly, we can give the general solutions of the left- and right-handed KK gravitinos without their boundary behaviors, then consider their boundary conditions at $y=0,\pi R$, respectively. Therefore, neglecting the term of $\sigma''$ and performing the transformations $z_n=\frac{m_n}{k}e^\sigma$, $\xi_{Ln,Rn}(y)=e^\sigma\bar{\xi}_{Ln,Rn}(z_n)$, the equation (\ref{CEOM}) will be transformed into the general $\nu-$order Bessel equation:
\begin{eqnarray}
z^2_n\frac{d^2}{dz_n^2}\bar{\xi}_{Ln,Rn}
+\left[z^2_n-\frac{1}{k^2}
\left(\frac{9}{4}\pm\frac{3}{2}\right)
\sigma'^{2}\right]\bar{\xi}_{Ln,Rn}=0.~~~~\label{TEOM}
\end{eqnarray}
The general solution is $\bar{\xi}_{Ln,Rn}=z_n^\alpha Z_{\nu_{\pm}}(\lambda z_n^\beta)$,
where the parameters are $\alpha=\frac{1}{2},~\beta=\lambda=1,
~\nu_{\pm}=|\frac{3}{2}\pm\frac{1}{2}|$, and $Z_{\nu}(x)$ is the $\nu-$order column function: $Z_{\nu}(x)=J_\nu(x)+b_\nu Y_\nu(x)$. The corresponding $y-$dependent left- and right-handed KK gravitinos read
\begin{eqnarray}
\xi_{Ln}(y)&=&\frac{1}{N_n}e^{\frac{3}{2}\sigma}
             \left[J_2\left(\frac{m_n}{k}e^\sigma\right)
              +b(m_n)Y_2\left(\frac{m_n}{k}e^\sigma\right)\right],\\
\xi_{Rn}(y)&=&\frac{1}{N_n}e^{\frac{3}{2}\sigma}
             \left[J_1\left(\frac{m_n}{k}e^\sigma\right)
              +b(m_n)Y_1\left(\frac{m_n}{k}e^\sigma\right)\right],~~~~~
\end{eqnarray}
where the coefficients $N_n$ and $b(m_n)$ are arbitrary constants. These constants are the same for left- and right-handed KK modes since they are obeying Eq.~(\ref{CEOM}), which is consistent with the result in \cite{Gherghetta2001}.

\begin{figure*}[htb]
\begin{center}
\includegraphics[width=7cm]{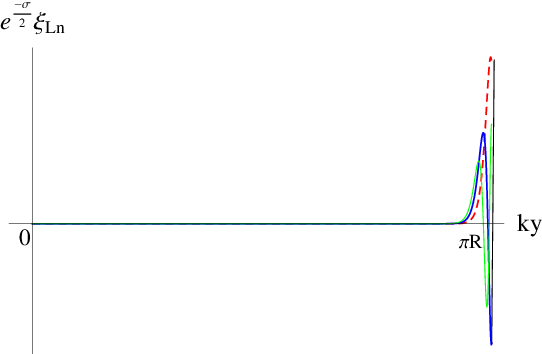}~~~~
\includegraphics[width=7cm]{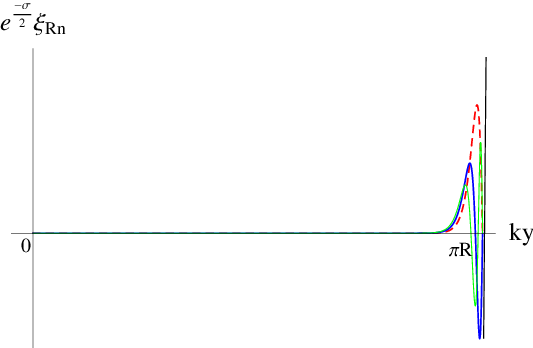}\\
\includegraphics[width=7cm]{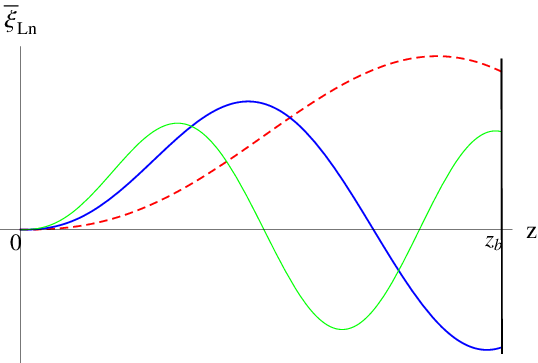}~~~~
\includegraphics[width=7cm]{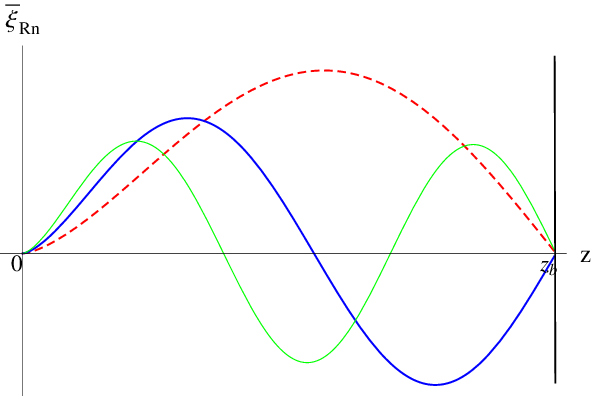}
\end{center}
\caption{\label{rsy1}The shapes of three lowest chiral massive KK modes in the RS1 model with the physical coordinate, $e^{-\frac{\sigma}{2}}\xi_{Ln,Rn}(y)$, and with the conformal coordinate, $\bar\xi_{Ln,Rn}(z)$, for the boundary conditions (\ref{RSBCS1}). The corresponding mass spectrum is given by $m_1=1.62508~\textrm{TeV}$ (the dashed red lines), $m_2=2.97546~\textrm{TeV}$ (the thickness blue lines), and $m_3=4.31474~\textrm{TeV}$ (the thin green lines). }
\end{figure*}

In the following, we focus on the boundary conditions imposing on $\xi_{Ln,Rn}$ at the fixed points $y=0,\pi R$ and then give the mass spectrum of the four-dimensional KK gravitinos.
It is clear that $\xi_{Ln,Rn}$ must be even (odd) as $\xi_{Rn,Ln}$ is odd (even) under the $Z_2$ symmetry,
since the operator $\left[\pm\partial_y+\left(\frac{3}{2}\mp1\right)\sigma'\right]$ in Eq. (\ref{CEOM}) is odd.
Firstly, considering that $\xi_{Ln}$ is even, while $\xi_{Rn}$ odd, we have
\begin{eqnarray}
\left(\partial_y+\frac{\sigma'}{2}\right)\xi_{Ln}\mid_{0,\pi R}=0,~~~~\xi_{Rn}\mid_{0,\pi R}=0.\label{RSBCS1}
\end{eqnarray}
From the above boundary conditions, $b(m_n)=-\frac{J_1(\frac{m_n}{k})}{Y_1(\frac{m_n}{k})}$, and the mass spectrum is determined by
\begin{eqnarray}
b(m_n)=b\left(m_ne^{\pi k R}\right)\label{Mass}.
\end{eqnarray}
In the limit $m_n\ll k \sim M_{Pl}$ (the light KK modes compared with Planck mass) and 
${m_n}\gg {k}e^{-k\pi R} \sim 1 $Tev (the KK modes with not too light mass compared with Tev scale), we obtain $b(m_n)\simeq\frac{m_n^2\pi}{4k^2}\rightarrow0$, and then
\begin{eqnarray}
J_1\left(\frac{m_n}{k}e^{k\pi R}\right)
    \simeq\sqrt{\frac{2ke^{-k\pi R}}{\pi m_n}}~
     \cos\left(\frac{m_n}{k}e^{k\pi R}-\frac{3\pi}{4}\right)\rightarrow0.\nonumber
\end{eqnarray}
Therefore, the KK masses can be obtained approximately as
\begin{eqnarray}
m_n\simeq\left(n+\frac{1}{4}\right)\pi ke^{-kR\pi}
\sim  3\left(n+\frac{1}{4}\right) {\textrm{TeV}}
\end{eqnarray}
with $n=1,~2,~3,\cdots$.

\begin{figure*}[htb]
\begin{center}
\includegraphics[width=7cm]{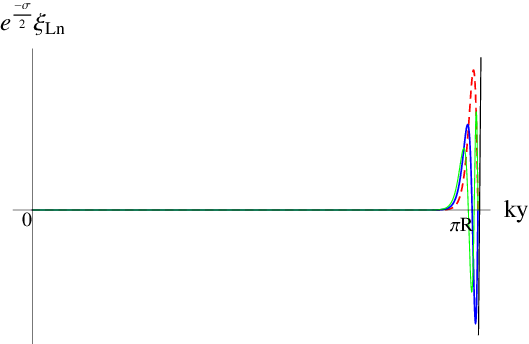}~~~~
\includegraphics[width=7cm]{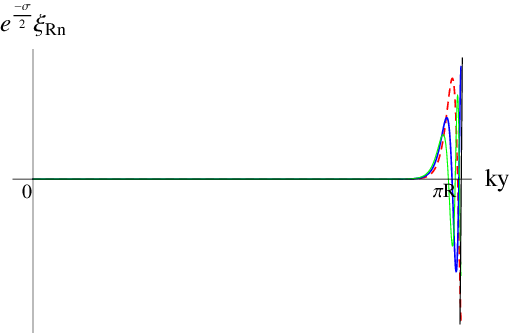}\\
\includegraphics[width=7cm]{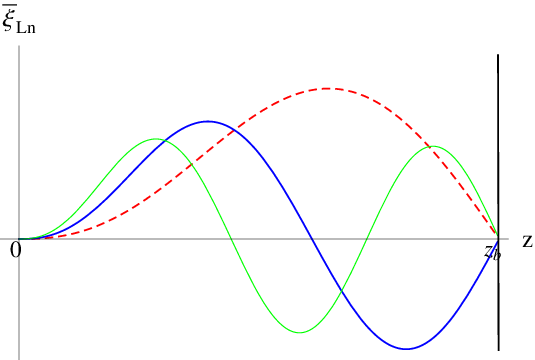}~~~~
\includegraphics[width=7cm]{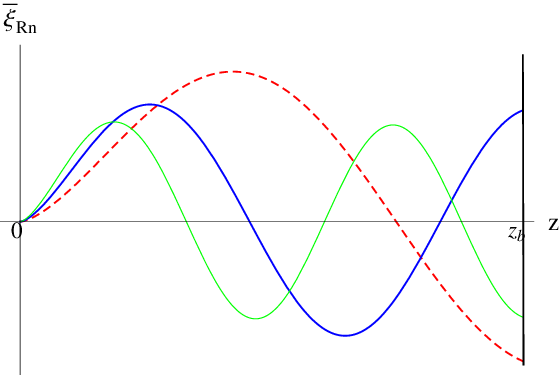}
\end{center}
\caption{\label{rsy2}The shapes of three lowest chiral massive KK modes in the RS1 model with the physical coordinate, $e^{-\frac{\sigma}{2}}\xi_{Ln,Rn}(y)$, and with the conformal coordinate, $\bar\xi_{Ln,Rn}(z)$, for the boundary conditions (\ref{RSBCS2}). The corresponding mass spectrum is given by $m_1=2.17811~\textrm{TeV}$ (the dashed red lines), $m_2=3.56988~\textrm{TeV}$ (the thickness blue lines), and $m_3=4.92815~\textrm{TeV}$ (the thin green lines). }
\end{figure*}

Secondly,
considering that $\xi_{Rn}$ is even and $\xi_{Ln}$ odd, we have the following boundary conditions
\begin{eqnarray}
\left(-\partial_y+\frac{5\sigma'}{2}\right)\xi_{Rn}\mid_{0,\pi R}=0,~~~~
\xi_{Ln}\mid_{0,\pi R}=0. \label{RSBCS2}
\end{eqnarray}
So, the coefficient $b(m_n)$ is given as
\begin{eqnarray}
b(m_n)=-\frac{J_2\left(\frac{m_n}{k}\right)}{Y_2\left(\frac{m_n}{k}\right)},
\end{eqnarray}
which satisfies the expression $b(m_n)=b\left(m_ne^{\pi k R}\right)$. In the limit $m_n\ll k$ and ${m_n}\gg {k}e^{-k\pi R} \sim 1 $Tev,  we have $b(m_n)\simeq\frac{\pi}{2}\left(\frac{m_n}{2k}\right)^4\rightarrow0$, and
\begin{eqnarray}
J_2\left(\frac{m_n}{k}e^{k\pi R}\right)
    \simeq\sqrt{\frac{2ke^{-k\pi R}}{\pi m_n}}~
     \cos\left(\frac{m_n}{k}e^{k\pi R}-\frac{5\pi}{4}\right)\rightarrow0.\nonumber
\end{eqnarray}
The KK masses are given approximately as
\begin{eqnarray}
m_n\simeq\left(n+\frac{3}{4}\right)\pi ke^{-kR\pi}
\sim  3\left(n+\frac{3}{4}\right) {\textrm{TeV}}.
\end{eqnarray}

Three lowest chiral massive KK modes in the coordinates $y$ and $z$ are plotted in Fig.~\ref{rsy1} (with the boundary conditions (\ref{RSBCS1})) and Fig.~\ref{rsy2} (with the boundary conditions (\ref{RSBCS2})). From the point of view of the conformal coordinate $z$, the lower KK modes are not localized near any brane. However, as the statement in Ref. \cite{Gherghetta2000}, from the point of view of the physical coordinate $y$, both the left- and right-handed lower KK modes are localized near the IR brane. {The different localization behaviours of these massive chiral KK modes with the conformal coordinate $y$ and the physical coordinate $z$ are induced by the coordinate transformation $dz=e^{\sigma}dy$. Taking a five-dimensional free vector field in the RS2 model as an example, we have known that the extra dimensional part of the vector zero mode is a constant from the physical coordinate of view, i.e., it is evenly distributed along the extra dimension $y$, while the extra dimensional part of the vector zero mode with the conformal coordinate $z$ is a function about the warp factor and it is localized on the brane at $z=0$.  In fact the physical localization property of these KK modes should be described exactly by the inherent distance, i.e., the physical coordinate $y$. }

In this section, we have reviewed the concrete process to calculate the chiral zero modes and a set of massive KK modes of the gravitino field. There is only one of the localized chiral zero modes, either the left-handed zero mode near the UV brane or the right-handed one near the IR brane. Both the lower left- and right-handed massive KK modes are confined to the IR brane from the point of view of the physical coordinate $y$. Here, we can see that even though two 3-branes are separated by a Planck scale distance $\pi R \sim 12 l_{pl}$, the mass gap of the KK modes is of about $\pi k e^{-kR\pi}\sim 3\textrm{TeV}$, which is very different from the KK theory \cite{Kaluza1921} and ADD theory \cite{Arkani-Hamed1998}. Furthermore, the left- and right-handed massive KK modes share the same mass spectrum but with different parity, and the lower ones are both localized near the IR brane.

\section{Localization of the gravitino field on the thin branes in the scalar-tensor theory}
\label{sec4}

Another typical thin brane model is the one in the scalar-tensor gravity theory \cite{Yang2012}. In this model, our world is confined to the positive tension brane rather than to the negative one in order to solve the gauge hierarchy problem, and the spacings of mass spectra for graviton and matter fields are very tiny (about $10^{-3}$eV) \cite{Yang2012,Xie2015}. In the following, we investigate the KK modes of the gravitino field on the scalar-tensor brane.

The action of this brane system with a dilaton field nonminimally coupled to gravity reads \cite{Yang2012}
\begin{eqnarray}
S_5=\frac{M_5^3}{2}\int dx^5 \sqrt{-g}~e^{\lambda\phi}\left[\mathcal{R}
-(3+4\lambda)\partial_M\phi~\partial^M\phi\right],~~~\label{Sscale-tensor}
\end{eqnarray}
where $\mathcal{R}$ is the five-dimensional scalar curvature, $M_5$ and $\phi$ are the five-dimensional scale of gravity and the dilaton field, respectively. Here we also consider the static flat brane scenario with an $S^1/Z_2$ extra spatial dimension. The metric is assumed to be the one given in Eq. (\ref{LE}) with the physical extra dimension coordinate $y$, or
\begin{eqnarray}
ds^2=e^{-2\sigma}\big(\eta_{\mu\nu}dx^\mu dx^\nu+dz^2\big) \label{LE2}
\end{eqnarray}
with the conformal extra dimension one $z\in[-z_b,z_b]$. Hence, the dilaton field $\phi$ can be supposed to be a function of the extra coordinate $z$, i.e., $\phi=\phi(z)$.  {Combined with Eq. (\ref{LE2}), the field equations derived from the action (\ref{Sscale-tensor}) are given as
\begin{eqnarray}
\lambda\ddot\phi+(\lambda^2+2\lambda+3/2)\dot\phi^2-2\lambda\dot\sigma\dot\phi
+3(\dot\sigma^2-\ddot\sigma)&=&0,\label{eq1}\\
(\dot\phi+2\dot\sigma)[(4\lambda+3)\dot\phi-6\dot\sigma]&=&0,\label{eq2}\\
(4\lambda+3)\ddot\phi+(2\lambda^2+3/2\lambda)\dot\phi^2
-3(4\lambda+3)\dot\sigma\dot\phi
-2\lambda(3\dot\sigma^2-2\ddot\sigma)&=&0,\label{eq3}
\end{eqnarray}
where the dot denotes the derivative with respect to the conformal coordinate $z$.}
There are two thin branes in this system, which are located at $z=0$ (IR brane) and $z_b$ (UV brane), contrary to the RS1 model. Furthermore, another difference from the RS1 model is that the five-dimensional fundamental scale is the electroweak scale, i.e.,  $M_5=M_{\textrm{ew}} \sim 1\textrm{TeV}$. {There are two kinds of brane solutions with different ranges of the coupling parameter $\lambda$, which are necessary to generate the brane configuration. These two kinds of brane solutions are physically equivalent from the view point of the four-dimensional graviton KK mass spectrum on the brane \cite{Yang2012}, which means that the four-dimensional graviton KK mass spectrum are the same for these two different solutions. However, the KK mass spectra of other four-dimensional matters (such as scale field, fermions, Kalb-Ramond field) are different for these two brane solutions \cite{Xie2015}. Next we will exhibit that the gravitino KK spectrum for two brane solutions are also not the same. From this point, we can say these two brane solutions are not physically equivalent.}

\begin{figure*}[htb]
\begin{center}
\includegraphics[width=7cm]{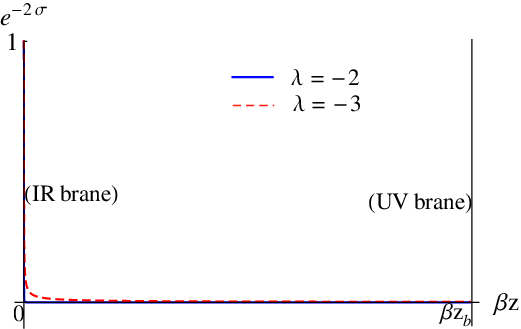}
\includegraphics[width=7cm]{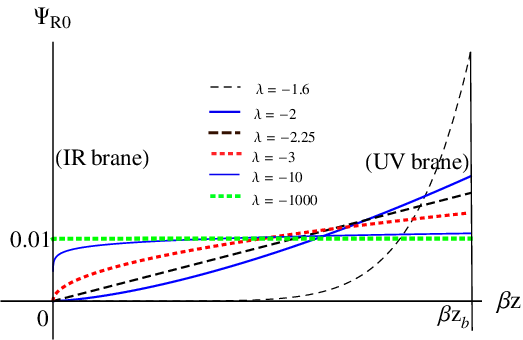}
\end{center}
\vskip -4mm \caption{The shapes of the warped factor $e^{-2\sigma}$ and the right-handed zero mode $\Psi_{R0}\equiv {\bar\xi_{R0}}/{\sqrt{\beta}}$ in the conformal coordinate $z$ for the first brane solution (\ref{Factor1}) in the scalar-tensor theory.}\label{WapedST1b}
\end{figure*}

\begin{figure*}[htb]
\begin{center}
\includegraphics[width=7cm]{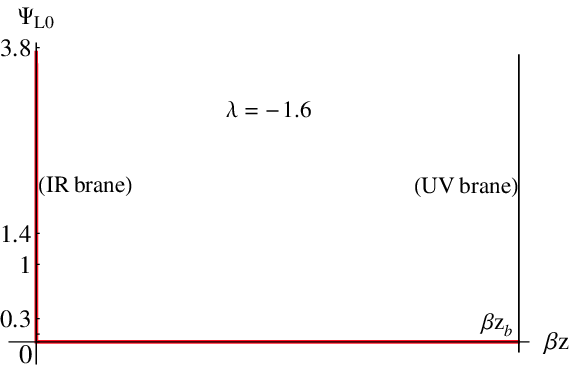}~~~~
\includegraphics[width=7cm]{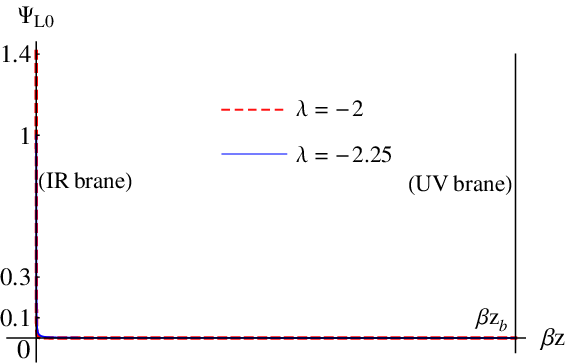}\\
\includegraphics[width=7cm]{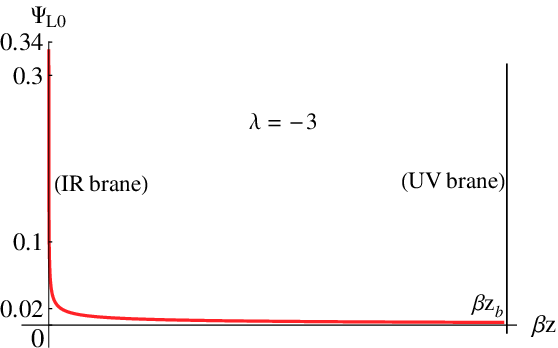}~~~~
\includegraphics[width=7cm]{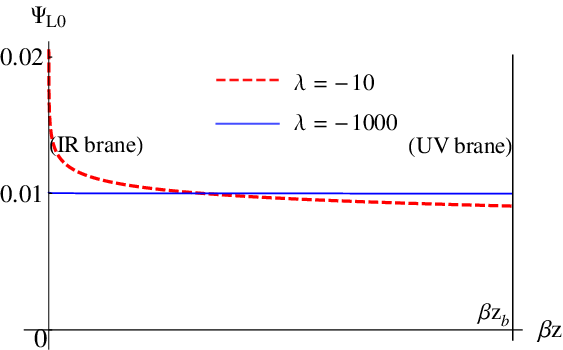}
\end{center}
\vskip -4mm \caption{The shapes of the left-handed zero mode $\Psi_{L0}\equiv {\bar\xi_{L0}}/{\sqrt{\beta}}$  in the conformal coordinate $z$  for the first solution (\ref{Factor1})  in the scalar-tensor theory. }\label{figFZ}
\end{figure*}

\begin{figure*}[htb]
\begin{center}
\includegraphics[width=7cm]{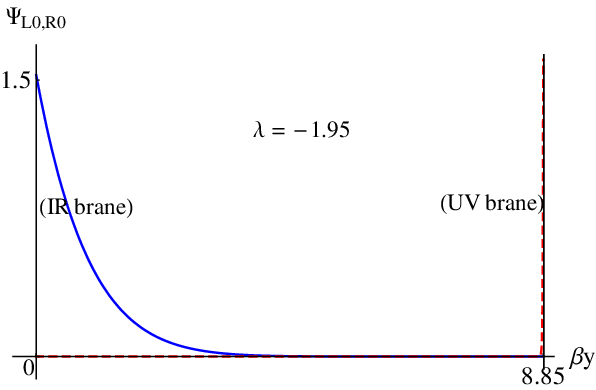}\\
\includegraphics[width=6.5cm]{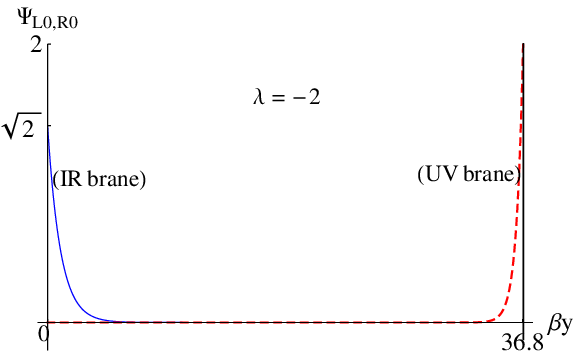}
\includegraphics[width=6.5cm]{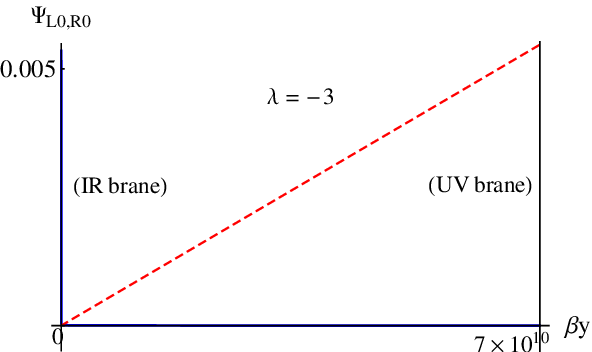}
\includegraphics[width=6.5cm]{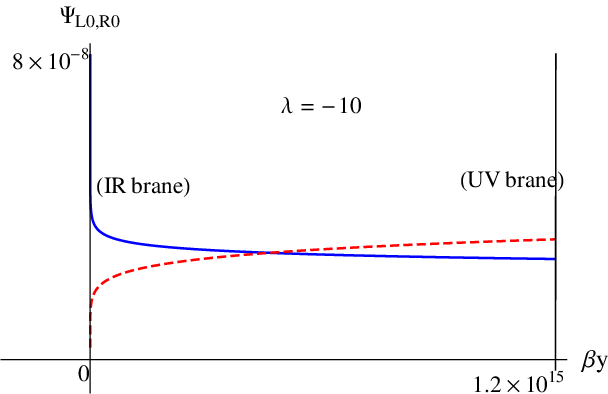}
\includegraphics[width=6.5cm]{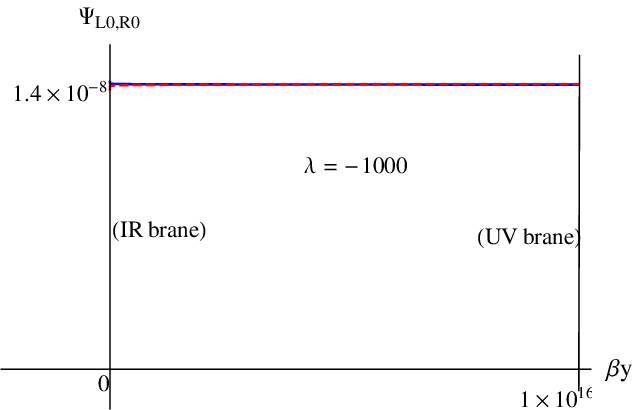}
\end{center}
\vskip -4mm \caption{The shapes of the chiral zero modes  $\Psi_{L0}\equiv {e^{-\frac{\sigma}{2}}\xi_{L0}}/{\sqrt{\beta}}$ (the blue  lines) and $\Psi_{R0}\equiv {e^{-\frac{\sigma}{2}}\xi_{R0}}/{\sqrt{\beta}}$ (the red dashed  lines) in the physical coordinate $y$
for the first solution (\ref{Factor1})  in the scalar-tensor theory.
}\label{figFZ1}
\end{figure*}

\subsection{Solution 1}

{Firstly, from the first bracket in the left of Eq. (\ref{eq2}), we have
\begin{eqnarray}
\dot\phi=-2\dot\sigma,
\end{eqnarray}
then substituting it into Eq. (\ref{eq1}), we obtain
\begin{eqnarray}
\ddot\sigma=(2\lambda+3)\dot\sigma^{2},~~~~
(\lambda\neq-3/2)
\end{eqnarray}
Thus the solution of the warped factor reads}
\begin{eqnarray}
e^{-2\sigma}=(1+\beta|z|)^{\frac{2}{3+2\lambda}},\label{Factor1}
\end{eqnarray}
where the parameters $\beta>0,~\lambda<-3/2$.  Its profile is plotted in Fig. \ref{WapedST1b}, which is similar with the RS1 model. For convenience, we redefine the parameters $p_1\equiv\frac{1}{3+2\lambda}(<0)$ and $\tilde{z}_b\equiv1+\beta z_b(>1)$. Then the physical coordinate $y$ is given by
\begin{eqnarray}
y&=&\int_0^z e^{-\sigma(w)}d w \nonumber \\
&=&\left\{\begin{array}{ll}
  \frac{1}{\beta(1+p_1)}\left[(1+\beta|z|)^{1+p_1}-1\right]~~~~~~~
  \lambda\neq-2\\
  \frac{1}{\beta} \ln(1+\beta|z|)~~~~~~~~~~~~~~~~~~~~~~~~~~\lambda=-2
\end{array}\right.,
\end{eqnarray}
from which we can see that the bulk size is not dynamically fixed and there should be the Goldberger-Wise mechanism \cite{Goldberger1999} to stabilize the size of the extra dimension. In particular, for $\lambda=-2$ the solution of the warp factor is the same as that in the RS1 model except that the energy scales $k$ and $\beta$ are different. Therefore, in the scalar-tensor brane model with  $\lambda=-2$ the massless and massive modes of the gravitino field are the same with that obtained in the RS1 model formally, and the only difference is the mass gap.

The solutions of the left- and right-handed zero modes $\bar\xi_{L0,R0}$ for the brane solution (\ref{Factor1}) are
\begin{eqnarray}
\bar\xi_{L0}(z)&=&\frac{1}{\sqrt{L_0}}\left(1+\beta|z|\right)^{\frac{3}{2}p_1},\\
\bar\xi_{R0}(z)&=&\frac{1}{\sqrt{R_0}}\left(1+\beta|z|\right)^{-\frac{3}{2}p_1}.
\end{eqnarray}
From the orthogonality conditions (\ref{orthogonality}), we have the normalized constants
\begin{eqnarray}
L_0&=&\left\{\begin{array}{ll}
\frac{1}{\beta}\ln\tilde{z}_b~~~~~~~~~~~~~~~~~~~~~~~~~~\lambda=-3\\
\frac{1}{(3p_1+1)\beta}\left(\tilde{z}_b^{3p_1+1}-1\right)
~~~~~~~\lambda\neq-3
\end{array}\right.,~~~~~~\\
R_0&=&\left\{\begin{array}{ll}
\frac{1}{2\beta}\tilde{z}_b^2~~~~~~~~~~~~~~~~~~~~~~~~~~~~\lambda=-3\\
\frac{1}{(1-3p_1)\beta}\left(\tilde{z}_b^{1-3p_1}-1\right)
~~~~~~~\lambda\neq-3
\end{array}\right..
\end{eqnarray}

The profiles of $\frac{\bar\xi_{L0,R0}(z)}{\sqrt{\beta}}$ and $\frac{e^{-\frac{\sigma}{2}}\xi_{L0,R0}(y)}{\sqrt{\beta}}$ with different values of $\lambda$ ($\lambda<-\frac{3}{2}$) are shown in Figs. \ref{WapedST1b}, \ref{figFZ}, and \ref{figFZ1}, from which we can see that there are some interesting characters for the chiral zero modes:
\begin{itemize}
 \item{In the conformal coordinate:
  \begin{itemize}
   \item{The right-handed zero mode is almost zero near the location of the IR brane, and then sharply increases toward the UV brane when $\lambda$ is closed to $-3/2$.
       This situation will happen for $-2.25<\lambda<-\frac{3}{2}$, and the growth is linear as $\lambda=-2.25$ (see {Fig. \ref{WapedST1b}}), which
       means that this mode is localized near the UV brane when the coupling parameter satisfies $-2.25\leq\lambda<-\frac{3}{2}$. With the continue decreasing of {$\lambda<-50$}, it will increase rapidly from the IR brane and rise slowly to the UV brane. For $\lambda \leq -1000$, the right-handed zero mode  is distributed uniformly along the extra dimension, which is the same as the left-handed one.}
   \item{The left-handed zero mode sharply decreases to zero with the increase of $\beta z$ (see Fig. \ref{figFZ}). This decreasing becomes slower and its value at $z=0$ decreases with the increase of $|\lambda|$ but stops after $\lambda \leq -50$. So the left-handed zero mode is localized near the IR brane at $z=0$ when the coupling parameter $-50 \leq \lambda < -\frac{3}{2}$, and
       it is distributed uniformly along the extra dimension when $\lambda \leq -1000$.}
 \end{itemize}}
 \item{In the physical coordinate:
 \begin{itemize}
  \item{The localization characters of the chiral zero modes are similar to the case in the conformal coordinate.}
 \end{itemize}}
\end{itemize}

The result is that, under the boundary condition $\bar\xi_{Ln}(z)|_{0,z_b}=0$, there is  one chiral zero mode localized on the IR brane only when $-50\leq\lambda\leq-\frac{3}{2}$, which is similar to the case of the RS1 brane model. However, both the left- and right-hand zero modes become a constant when $\lambda \leq -1000$.

\begin{figure*}[htb]
\begin{center}
\includegraphics[width=11cm]{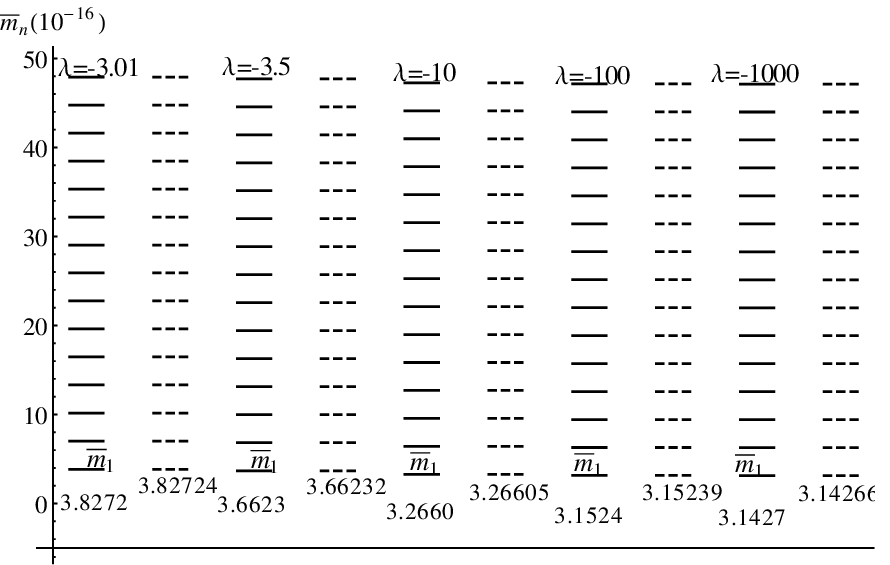}
\end{center}
\vskip -4mm \caption{ The mass spectra of the gravitino field with the boundary condition $\bar\xi_{Rn}(\bar z)|_{0,\bar z_b}=0$ for the first solution (\ref{Factor1})  in the scalar-tensor theory. The solid and dashed lines stand for the accurate (numerical) and approximate (analytical) mass spectra, respectively. The parameters are set to $|\bar z_b|=10^{16}$, and $\lambda=-3.01,-3.5,-10,-100,-10000$.}\label{ML1ST}
\end{figure*}

\begin{figure*}[htb]
\begin{center}
\includegraphics[width=7cm]{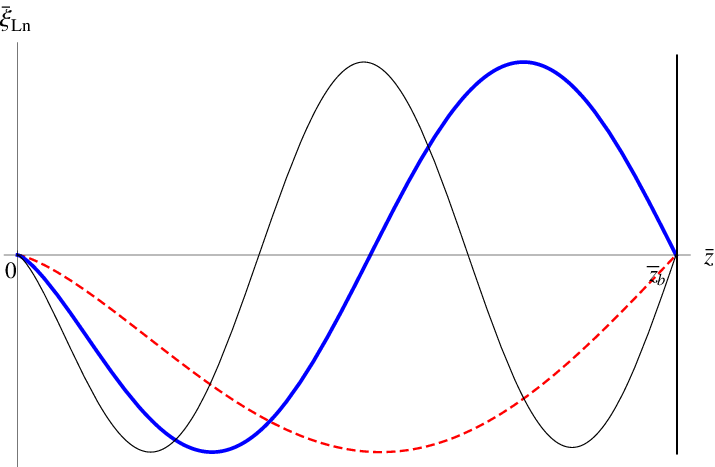}~~~~
\includegraphics[width=7cm]{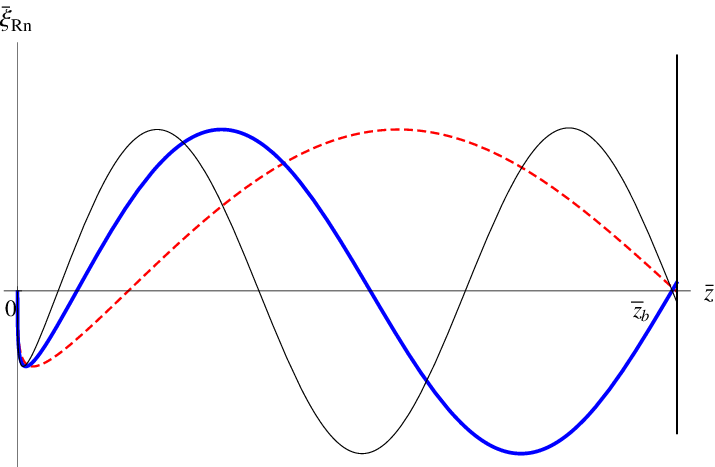}
\end{center}
\vskip -4mm \caption{ The shapes of three lowest chiral massive KK modes $\bar\xi_{Ln,Rn}(\bar z)$ with the boundary condition $\bar\xi_{Rn}(\bar z)|_{0,\bar z_b}=0$  for the first solution (\ref{Factor1})  in the scalar-tensor theory. The parameter $\lambda$ is set to $-3.5$. The corresponding mass spectrum is $\bar m_1=3.66232\times10^{-16}$ (the red dashed lines), $\bar m_2=6.83487\times10^{-16}$ (the blue thickness lines), $\bar m_3=9.98813\times10^{-16}$ (the black thin lines).  }\label{Wavefunction1L}
\end{figure*}

{In order to investigate the gravitino massive KK modes, we redefine the following dimensionless parameters:
\begin{eqnarray}
\bar {|z|}\equiv\beta|z|,~~~~
\bar {m}_n\equiv\frac{m_n}{\beta},~~~~
\bar {V}^{(1)}_{L,R}(\bar z)\equiv\frac{V^{(1)}_{L,R}(z)}{\beta^2},\label{refines}
\end{eqnarray}
so the Schr\"{o}dinger-like equations (\ref{Seq}) can be written as
\begin{eqnarray}
\left(-\partial_{\bar z}^2+\bar V^{(1)}_{L,R}(\bar z)\right)\bar{\xi}_{Ln,Rn}(\bar z)=\bar m_n^2\bar{\xi}_{Ln,Rn}(\bar z)
\end{eqnarray}
with the corresponding effective potentials
\begin{eqnarray}
\bar {V}^{(1)}_{L,R}(\bar z)=\frac{3(3p_1\mp2)p_1}{4(1+|\bar z|)^2}
      \pm\frac{3 p_1[\delta(\bar z)-\delta(\bar z-\bar z_b)]}{{(1+|\bar z|)}}. ~~~\label{V11}
\end{eqnarray}
To obtian the general solution of the left- and right-handed gravitino KK modes, we solve the above Schr\"{o}dinger-like equations by neglecting the delta functions {in (\ref{V11})} firstly. Then considering the boundary conditions on the general solutions {of the} gravitino KK modes, we can obtain the chiral gravitino KK mass spectra. The general solutions of the left- and right-handed KK modes read
\begin{eqnarray}
\bar\xi_{Ln}(\bar z)
&=&\frac{1}{N_n}\bigg(M_{0,-\lambda p_1}(\bar z)
+c(\bar m_n)W_{0,-\lambda p_1}(\bar z)\bigg),~~~\label{L1}\\
\bar\xi_{Rn}(\bar z)
&=&\frac{1}{N_n}\bigg(M_{0,-p_2}(\bar z)+c(\bar m_n)W_{0,-p_2}(\bar z)\bigg),
 \label{R1}
\end{eqnarray}
where $M_{a,b}(\bar z)$ and $W_{a,b}(\bar z)$ are two kinds of Whittaker functions, and $p_2\equiv\frac{1+3p_1}{2}$.} Note that two constants $N_n$ and $c(\bar m_n)$ are the same for left- and right-handed KK modes since $\bar\xi_{Ln,Rn}$ satisfy the first-order equations (\ref{CEOMz}).
The boundary conditions (\ref{BCD}) will determine the constant $c(\bar m_n)$ and mass spectrum $\bar m_n$, while the orthonormal conditions (\ref{orthogonality}) will fix $N_n$.

\begin{figure*}[htb]
\begin{center}
\includegraphics[width=11cm]{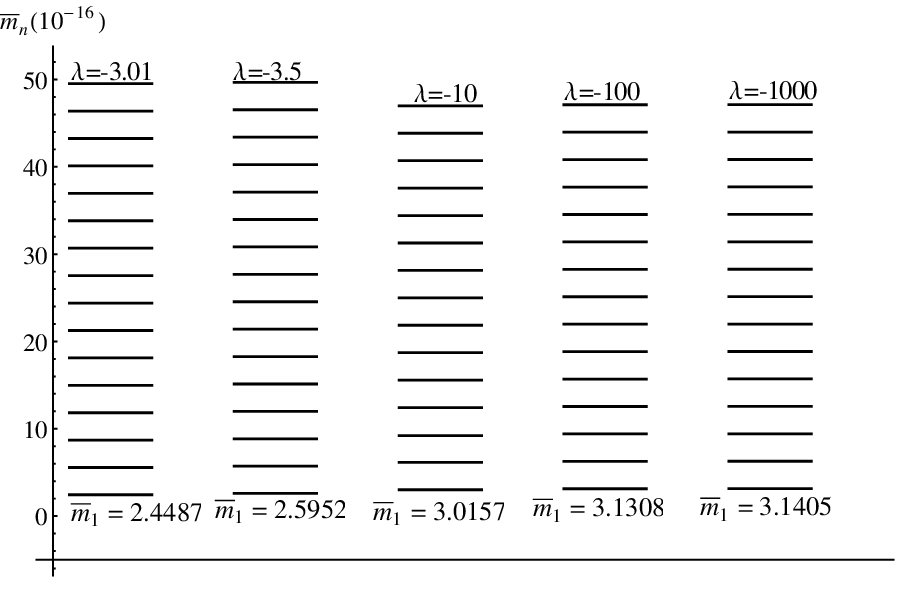}
\end{center}
\vskip -4mm \caption{The mass spectrum of the gravitino
field with the boundary condition $\bar\xi_{Rn}(\bar z)|_{0,\bar z_b}=0$
for the first solution (\ref{Factor1})  in the scalar-tensor theory.
The parameters are set to $|\bar z_b|=10^{16}$,
and $\lambda=-3.01,-3.5,-10,-100,-10000$.}
\label{MR1ST}
\end{figure*}

\begin{figure*}[htb]
\begin{center}
\includegraphics[width=7cm]{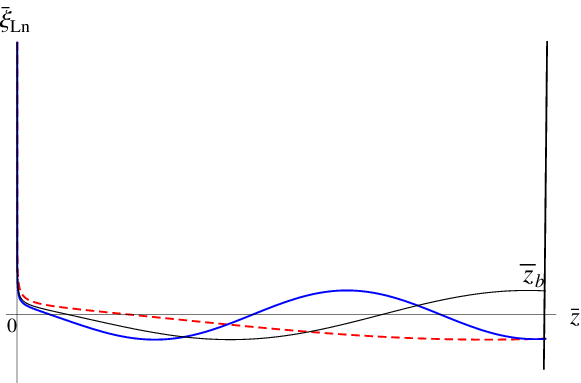}~~~~
\includegraphics[width=7cm]{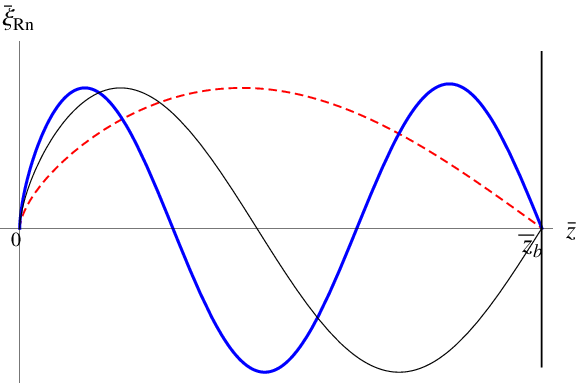}
\end{center}
\vskip -4mm \caption{The shapes of three lowest chiral massive KK modes $\bar\xi_{Ln,Rn}$ with the boundary condition $\bar\xi_{Rn}(\bar z)|_{0,\bar z_b}=0$  for the first solution (\ref{Factor1}) in the scalar-tensor theory. The parameter $\lambda$ is set to $-3.5$. The corresponding mass spectrum is $\bar m_1=2.59517\times10^{-16}$ (the red dashed lines), $\bar m_2=5.7144\times10^{-16}$ (the blue thickness lines), $\bar m_3=8.84895\times10^{-16}$ (the black thin lines).  }
\label{Wavefunction1R}
\end{figure*}

For the boundary conditions (\ref{BCD}), we have two choices for the left- and right-handed massive KK modes: $\bar\xi_{Ln}(\bar z)|_{0,\bar z_b}=0$, or $\bar\xi_{Rn}(\bar z)|_{0,\bar z_b}=0$. We consider the first boundary condition $\bar\xi_{Rn}(\bar z)|_{0,\bar z_b}=0$, i.e., the left-handed KK mode satisfies Dirichlet boundary condition, for which the boundary condition of right-handed KK mode should be $\partial_z\bar\xi_{Ln}(\bar z)\mid_{0}
=\frac{3p_1}{\beta}\bar\xi_{Ln}(0)$,and $\partial_z\bar\xi_{Ln}(\bar z)\mid_{\bar z_b}=-\frac{3p_1}{\beta(1+|\bar z_b|)}\bar\xi_{Ln}(\bar z_b)$. The corresponding mass spectrum $\bar m_n$ is determined by
\begin{eqnarray}
\frac{M_{0,-\lambda p_1}(2\textrm i~\bar m_n)}{W_{0,-\lambda p_1}(2\textrm i~\bar m_n)}
=\frac{M_{0,-\lambda p_1}(2\textrm i~\bar m_n[1+|\bar z_b|])}
  {W_{0,-\lambda p_1}(2\textrm i~\bar m_n[1+|\bar z_b|])}. \label{MSL1}
\end{eqnarray}
In the limit $\bar m_n\ll1$ and  $1\ll\bar z_b$, the left side of (\ref{MSL1}) is approximate to zero and the four-dimensional KK mass spectrum $\bar m_n$ is given by
\begin{eqnarray}
\bar m_n\simeq \frac{x_n}{2(1+|\bar z_b|)}\simeq \frac{x_n}{2|\bar z_b|},~~~~~(n=1,2,\cdots)
\end{eqnarray}
where $x_n$ satisfies $M_{0,-\lambda p_1}(ix_n)=0$. The accurate and approximate mass spectra of the gravitino field with the coupling parameter $\lambda$ are shown in Fig. \ref{ML1ST}, which show that the approximate mass spectrum in this long wave limit is nearly close to the numerical result. The mass of the lightest massive KK mode decreases with the increase of $|\lambda|$. The mass gap of two adjoining KK modes ($\Delta \bar m_n\simeq\frac{x_{n+1}-x_n}{2}10^{-16}$) decreases with the level $n$ and it will not change for heavy KK modes. And it also becomes smaller as decreasing the nonminimal coupling parameter $\lambda$.

For the second boundary condition $\bar\xi_{Ln}(\bar z)|_{0,\bar z_b}=0$, it will result in $\partial_{\bar z}\bar\xi_{Rn}(\bar z)\mid_{0}
=-\frac{3p_1}{\beta}\bar\xi_{Rn}(0)$, $\partial_{\bar z}\bar\xi_{Rn}(\bar z)\mid_{\bar z_b}
=\frac{3p_1}{\beta(1+|\bar z_b|)}\bar\xi_{Rn}(\bar z_b)$, and the mass spectrum $\bar m_n$ is determined by
\begin{eqnarray}
\frac{M_{0,-p_2}(2\textrm i~\bar m_n)}{W_{0,-p_2}(2\textrm i~\bar m_n)}
=\frac{M_{0,-p_2}(2\textrm i~\bar m_n[1+|\bar z_b|])}
  {W_{0,-p_2}(2\textrm i~\bar m_n[1+|\bar z_b|])}.\label{MSR1}
\end{eqnarray}
The mass spectrum in this brane solution is shown in Fig. \ref{MR1ST} for different values of $\lambda$. It can be seen that the mass of the lightest KK mode increases with $|\lambda|$, which is just opposite to the case of the first boundary condition $\bar\xi_{Ln}(\bar z)|_{0,\bar z_b}=0$. The mass gap of two adjoin KK masses ($\Delta \bar m_n$) changes with $n$ and $\lambda$ in the same way with the case of the boundary condition $\bar\xi_{Ln}(\bar z)|_{0,\bar z_b}=0$. Here, we only plot the left- and right-handed massive gravitino KK modes with $\lambda=-3.5$ for both two kinds of boundary conditions in Figs. \ref{Wavefunction1L} and \ref{Wavefunction1R}, respectively.

\subsection{Solution 2}

\begin{figure*}[htb]
\begin{center}
\includegraphics[width=7.5cm]{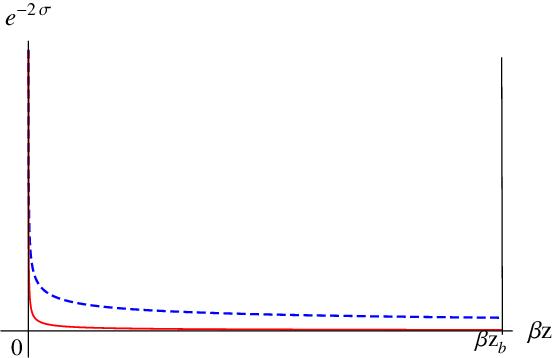}~~~~
\includegraphics[width=7.5cm]{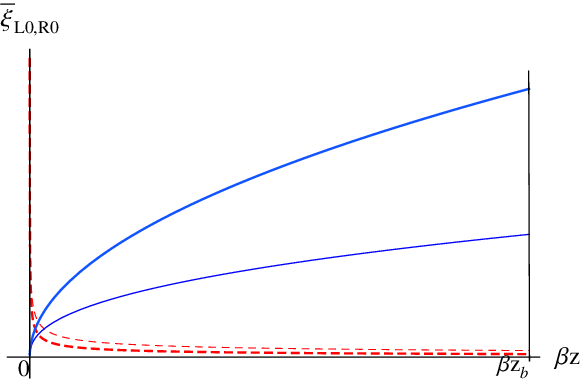}
\end{center}
\vskip -4mm \caption{The shapes of the warped factor $e^{-2\sigma}$ and the zero modes $\bar\xi_{L0}$ (the dashed lines), $\bar\xi_{R0}$ (the thick lines) for the second solution (\ref{Factor2})  in the scalar-tensor theory. The parameter $\lambda$ is set to $\lambda=-0.98$ for the thin lines, $\lambda=-1$ for the thick lines. }\label{STz}
\end{figure*}
\begin{figure*}[htb]
\begin{center}
\includegraphics[width=11cm]{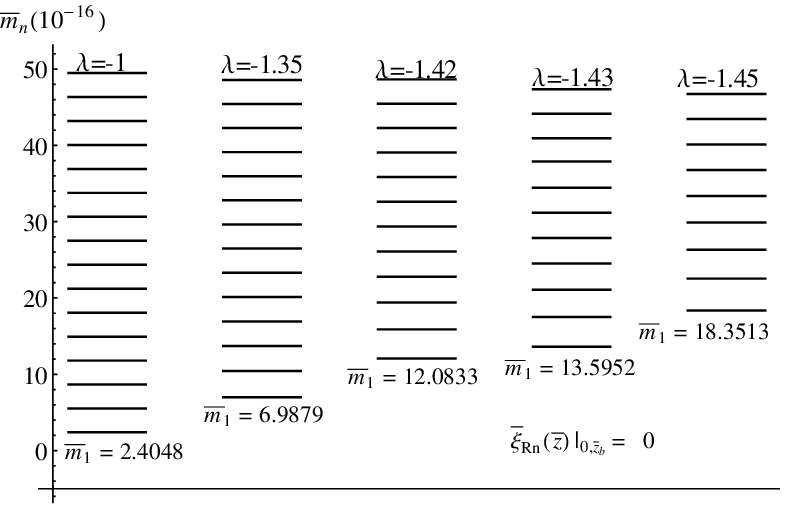}~~~~~\\
\includegraphics[width=11cm]{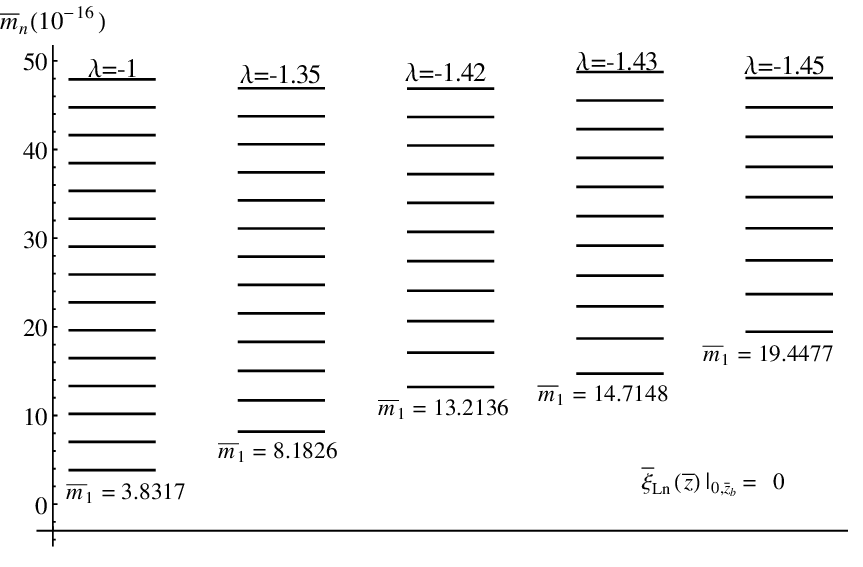}
\end{center}
\vskip -4mm \caption{The mass spectra of the gravitino field for two kinds of the boundary conditions in the scalar-tensor theory with the second solution (\ref{Factor2}). The parameters are set to $|\bar z|=10^{16}$, and $\lambda=-1,-1.35,-1.42,-1.43,-1.45$.}\label{M2}
\end{figure*}

{From the second bracket in the left of Eq. (\ref{eq2}), we have $\dot\sigma=\frac{3+4\lambda}{6}\dot\phi$. As considering this relation between $\dot\sigma$ and $\dot\phi$, field equations (\ref{eq1}) and (\ref{eq3}) are both rewritten as
\begin{eqnarray}
   \ddot\phi=\left(\frac{3}{2}+\lambda\right)\dot\phi^2,~~~~
   \left(\lambda\neq-\frac{3}{2}~~and~~
   \lambda\neq-\frac{3}{4}\right) \nonumber
\end{eqnarray}
Solving above equation, the solution of warped factor reads as in Ref.  \cite{Yang2012}}
\begin{eqnarray}
  e^{-2\sigma}=(1+\beta|z|)^{\frac{2(3+4\lambda)}{9+6\lambda}}.~~
 \left(-\frac{3}{2}<\lambda<-\frac{3}{4}\right) \label{Factor2}
\end{eqnarray}
With the redefined parameters (\ref{refines}), we have
\begin{eqnarray}
\bar y\equiv\beta y 
   =\frac{9+6 \lambda}{2(6+5\lambda)}
    \left[(1+|\bar z|)^{\frac{2}{3}(2-\lambda p_1)}
     -1\right],
\end{eqnarray}
and the chiral zero modes $\bar\xi_{L0,R0}(\bar z)$ read
\begin{eqnarray}
\bar\xi_{L0}(\bar z)
 &=&\frac{1}{\sqrt{L_0}}(1+|\bar z|)^{\frac{1}{2}(1+2\lambda p_1)},\\
\bar\xi_{R0}(\bar z)
 &=&\frac{1}{\sqrt{R_0}}(1+|\bar z|)^{-\frac{1}{2}(1+2\lambda p_1)}.
\end{eqnarray}
It is clear that the size of extra dimension is determined by the coupling parameter $\lambda$, which is similar to that in the first brane solution. The warp factor and chiral zero modes of the gravitino field are plotted in Fig. \ref{STz}. Note that as the statements above, there will be only one chiral zero mode localized near one brane as considering the boundary conditions (\ref{BCD}).

The effective potentials (\ref{effectiveV}) with the brane solution (\ref{Factor2}) read
\begin{eqnarray}\label{V2}
V^{(2)}_L(\bar z)&=&\frac{3(3p_1-2)p_1}{4(1+|\bar z|)^2}\nonumber\\
     && -\frac{(3p_1-2)[\delta(\bar z)-\delta(\bar z-\bar z_b)]}{\beta(1+|\bar z|)},\label{V2L}\\
V^{(2)}_R(\bar z)&=&\frac{(3p_1-2)(3p_1-4)}{4(1+|\bar z|)^2}\nonumber\\
     &&+\frac{(3p_1-2)[\delta(\bar z)-\delta(\bar z-\bar z_b)]}{\beta(1+|\bar z|)}.\label{V2R}
\end{eqnarray}
Note that the effective potentials (\ref{V11}) and (\ref{V2}) in the Shr\"{o}dinger-like equations for two kinds of brane solutions are the same as $p_1=1/3$, which means that two brane solutions stand for the same geometric configuration of spacetime.

\begin{figure*}[htb]
\begin{center}
\includegraphics[width=6.5cm]{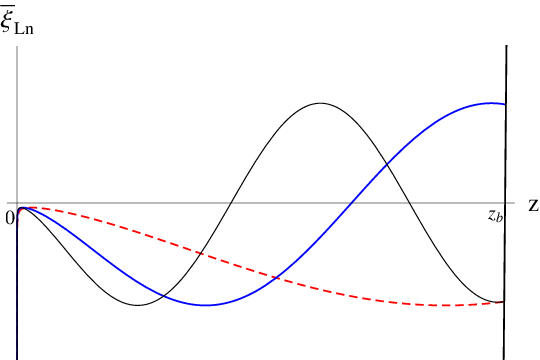}
\includegraphics[width=6.5cm]{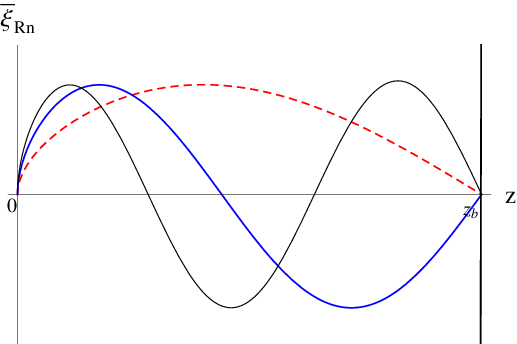}
\end{center}
\vskip -4mm \caption{The shapes of three lowest chiral massive KK modes $\bar\xi_{Ln,Rn}$ with the boundary condition $\bar\xi_{Rn}(\bar z)|_{0,\bar z_b}=0$ for the second solution (\ref{Factor2}) in the scalar-tensor theory. The parameter $\lambda$ is set to $-1$. The corresponding mass spectrum is $\bar m_1=2.4048\times10^{-16}$ (the red dashed lines), $\bar m_2=5.56445\times10^{-16}$ (the blue thickness lines), $\bar m_3=8.69877\times10^{-16}$ (the black thin lines). }\label{Wavefunction2R}
\end{figure*}

\begin{figure*}[htb]
\begin{center}
\includegraphics[width=6.5cm]{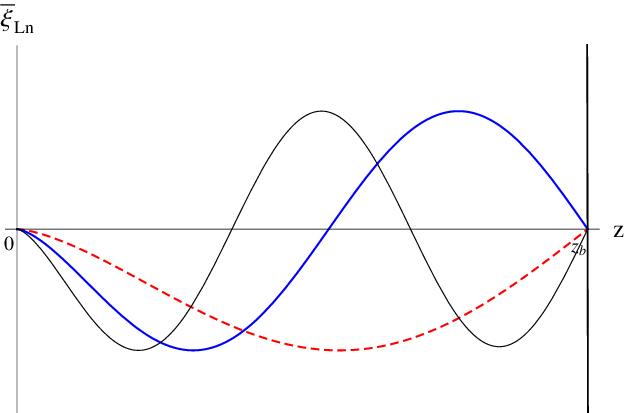}~~~~
\includegraphics[width=6.5cm]{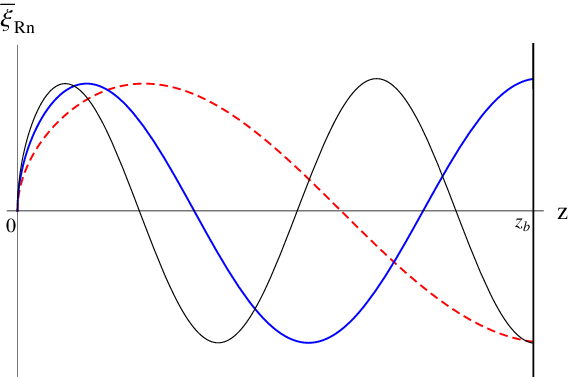}
\end{center}
\vskip -4mm \caption{The shapes of three lowest chiral massive KK modes $\bar\xi_{Ln,Rn}$ with the boundary condition $\bar\xi_{Ln}(\bar z)|_{0,\bar z_b}=0$ for the second solution (\ref{Factor2}) in the scalar-tensor theory. The parameter $\lambda$ is set to $-1$. The corresponding mass spectrum is $\bar m_1=3.83171\times10^{-16}$ (the red dashed lines), $\bar m_2=7.01559\times10^{-16}$ (the blue thickness lines), $\bar m_3=10.1735\times10^{-16}$ (the black thin lines). }\label{Wavefunction2L}
\end{figure*}

The general solutions of the left- and right-handed massive modes $\bar\xi_{Ln,Rn}(\bar z)$ are given by
\begin{eqnarray}
\bar\xi_{Ln}(\bar z)
\!\!&=&\!\!\frac{1}{N_n}\Big(M_{0,-\lambda p_1}(\bar z)
 +d(\bar m_n)W_{0,-\lambda p_1}(\bar z)\Big),\label{L2}\\
\bar\xi_{Rn}(\bar z)
\!\!&=&\!\!\frac{1}{N_n}\Big(M_{0,-1-\lambda p_1}(\bar z)
 +d(\bar m_n)W_{0,-1-\lambda p_1}(\bar z)\Big),~~~\label{R2}
\end{eqnarray}
where the constants $N_n$ and $d(\bar m_n)$ are the same for both the left- and right-handed KK modes. Performing the similar calculation with above subsection, the constant $N_n$ will be fixed by the orthonormal conditions (\ref{orthogonality}), and the mass spectrum $\bar m_n$ and $d(\bar m_n)$ will be determined by the boundary conditions (\ref{BCD}).

Firstly, we consider $\bar\xi_{Rn}(\bar z)|_{0,\bar z_b}=0$, i.e., the right-handed KK mode satisfies Dirichlet boundary condition, for which the left-handed KK mode at the boundaries satisfies $\partial_{\bar z}\bar\xi_{Ln}(\bar z)\mid_{0}
=-\frac{(2-3p_1)}{\beta}\bar\xi_{Ln}(0)$ and $\partial_{\bar z}\bar\xi_{Ln}(\bar z)\mid_{\bar z_b}
=\frac{(2-3p_1)}{\beta(1+|\bar z_b|)}\bar\xi_{Ln}(\bar z_b)$. The corresponding mass spectrum $\bar m_n$ is determined by the following equation
\begin{eqnarray}
\frac{M_{0,-\lambda p_1}(2\textrm i~\bar m_n)}{W_{0,-\lambda p_1}(2\textrm i~\bar m_n)}
=\frac{M_{0,-\lambda p_1}(2\textrm i~\bar m_n[1+|\bar z_b|])}
  {W_{0,-\lambda p_1}(2\textrm i~\bar m_n[1+|\bar z_b|])}.\label{MS2}
\end{eqnarray}
For the second boundary condition: $\bar\xi_{Ln}(\bar z)|_{0,\bar z_b}=0$, the right-handed KK modes at the boundaries satisfy $\partial_{\bar z}\bar\xi_{Rn}(\bar z)\mid_{0}
=\frac{(2-3p_1)}{\beta}\bar\xi_{Rn}(0)$, $\partial_{\bar z}\bar\xi_{Rn}(\bar z)\mid_{\bar z_b}
=-\frac{(2-3p_1)}{\beta(1+|\bar z_b|)}\bar\xi_{Rn}(\bar z_b)$, and the mass spectrum $\bar m_n$ is determined by
\begin{eqnarray}
\frac{M_{0,-(1+\lambda p_1)}(2\textrm i~\bar m_n)}{W_{0,-(1+\lambda p_1)}(2\textrm i~\bar m_n)}
=\frac{M_{0,-(1+\lambda p_1)}(2\textrm i~\bar m_n[1+|\bar z_b|])}
  {W_{0,-(1+\lambda p_1)}(2\textrm i~\bar m_n[1+|\bar z_b|])}.\label{MSL2}
\end{eqnarray}
The numerical result of the gravitino mass spectra for two kinds of boundary conditions is listed in Fig. \ref{M2}, which shows that the mass $\bar m_1$ of the lightest massive KK mode increases with the coupling parameter $|\lambda|$ for both two boundary conditions. Furthermore, the mass gap $\Delta \bar m_n$ of two adjacent KK massive modes becomes smaller with the decrease of $|\lambda|$ and it will not change for the heavy chiral KK modes. These results are not all the same with that in the first brane solution. From Figs.~\ref{ML1ST},~\ref{MR1ST}, and \ref{M2}, we know that these two brane solutions are not physically equivalent from the view of the gravitino mass spectrum. Here, we only show three lowest chiral massive KK modes with the conformal coordinate $z$ for different boundary conditions as $\lambda=-1$ in Figs. \ref{Wavefunction2R} and \ref{Wavefunction2L}, respectively. {Note that for both two brane solutions in the scalar-tensor model, the coordinate transformation $dz=e^\sigma dy$ do not change the localization property of lowest chiral massive KK modes.}

\section{Summary}\label{secConclusion}

In this paper, we have reviewed and investigated the massless and massive KK modes of a bulk massive gravitino field in the RS1 and scalar-tensor thin brane models, respectively. The localization and mass spectrum the gravitino field on a brane are obtained. Furthermore, we also compared the localization property of the gravitino field in these two thin brane models.

In the case of the RS1 model, the left- and right-handed zero modes (four-dimensional massless gravitinos) are localized near the Planck brane (UV brane at $y=0$) and Tev brane (IR brane at $y=y_b$), respectively. However, there is either the left- or right-hande massless gravitino survived as the result of the boundary conditions. The left- and right-handed massive KK modes share the same mass spectrum and the mass gap is of about 3TeV. The localization behaviors of the left- and right-handed lower massive KK modes in the conformal and physical coordinates are different. From the point of view of the conformal coordinate, the lower massive KK modes are not localized near any brane. However, they are confined near the IR brane from the point of view of the physical one. {This different results are induced by the coordinate transformation between the physical coordinate and the conformal one.} These behaviors of a gravitino field are similar to the Dirac fermion field in the RS1 model.

For the scalar-tensor thin brane model, there are two brane solutions.  The left- and right-handed zero modes are localized near the IR brane (at $y=0$ or $z=0$) and UV brane (at $y=y_b$ or $z=z_b$), respectively, if the coupling parameter $|\lambda|$ is not too large for the first brane solution and not close to $3/4$ for the second one. {For simplicity, we adopt the dimensionless parameter $\bar m_n$ to investigate the mass spectra of gravitino. We found that} the dimensionless mass spectra and localization properties of the gravitino field for two brane solutions are different. { {The} dimensionless mass gap is very small, about $10^{-16}$. {However,} there is no experimental signal of the gravitino at current. It may be due to the very weak interaction between the gravitino and other matters on the scalar-tensor thin brane, which can be be viewed as the reason why the gravitino can be regarded as one of the candidates of the dark matters. {If the coupling parameters of the KK gravitinos and other matters on a brane are about $1/M_{pl}$, just as the case of the KK gravitons, then these KK gravitinos would be invisible under TeV energy scale.}} For the first brane solution, the mass $\bar m_1$ of the lightest massive gravitino decreases with the coupling $|\lambda|$ for the boundary conditions $\bar\xi_{Ln}(\bar z)|_{0,\bar z_b}=0$, while it will increases with $|\lambda|$ for $\bar\xi_{Rn}(\bar z)|_{0,\bar z_b}=0$. For the second one, the mass $\bar m_1$ increases with the coupling $|\lambda|$ for both boundaries.
The lower massive KK modes are not localized near any brane except for the first brane solution with $\bar\xi_{Rn}(\bar z)|_{0,\bar z_b}=0$, for which the left-handed KK modes are confined near the IR brane. This is another difference from that in the RS1 model. These results show that the two scalar-tensor brane solutions are not physically equivalent, {although the mass spectra of the KK gravitons are the same.} From the above analysis, it is clear that the localization property of a bulk gravitino in the scalar-tensor thin brane model is richer than that in the RS1 model.

Furthermore, the linearized Brans-Dicke gravity can be recovered on either brane in the RS1 model with different Brans-Dicke parameters \cite{Garriga2000}. In the future work, we would like to consider the check by the solar system on the scalar-tensor brane model. We will also investigate interaction between a bulk gravitino and other fields.

\section*{Acknowledgments} \hspace{5mm}
We would like to thank Dr. Yuan Zhong for  helpful discussion. This work was supported by the National Natural Science Foundation of China (Grant No. 11522541, No. 11375075, and No. 10905027) and the Fundamental Research Funds for the Central Universities (Grant No. lzujbky-2016-k04). Y. Zhong was supported by the scholarship granted by the Chinese Scholarship Council (CSC).

\end{document}